\documentclass[prx,twocolumn, amsmath,amssymb,
reprint,%
author-year,%
author-numerical,%
dvipdfmx,
]{revtex4-1}

\usepackage{graphicx}
\usepackage{bm}
\usepackage{color}

\begin{document}

\title{Hidden layered structures from carbon-analog metastability in metal dichalcogenides}
\author{Shota Ono}
\email{shotaono@muroran-it.ac.jp}
\affiliation{Department of Sciences and Informatics, Muroran Institute of Technology, Muroran 050-8585, Japan}

\begin{abstract}
Carbon exhibits both a layered ground state structure that produces two-dimensional (2D) nanosheets and a non-layered diamond structure created under high pressure conditions. Motivated by this metastability relationship, we revisit the ground state structure of metal dichalcogenides that are known to have non-layered pyrite-type structure. Ultrathin films of pyrite-type ZnSe$_2$ spontaneously transform into a layered phase. This phase is identified as a ground state, and the monolayer exhibits strong elastic anisotropy and a semiconducting bandgap larger than that of the pyrite phase by a factor of two. We demonstrate that a two-valued but directional potential energy surface exists along a Bain-like distortion path, hiding the layered ground state. This work implies that many 2D materials are hidden in non-layered materials and connects 2D materials science with surface and high-pressure science. 
\end{abstract}

\maketitle

{\it Introduction---}Crystal structure prediction has a long history in materials science \cite{alberi2019}. Recent development of van der Waals (vdW) two-dimensional (2D) materials science requires an approach that accurately predicts the stable monolayers. A few thousands of 2D materials have been predicted \cite{lebegue,ashton,mounet,C2DB2021}, including vdW, non-vdW, and synthetic-types \cite{novoselov2020}. Nevertheless, only a few hundreds of them are actually synthesized \cite{X2DB}. The discovery of novel 2D materials is not only important in terms of their intrinsic properties \cite{burch2018,li2021} and potential applications \cite{ahn2020,schaibley2016}, but also enables us to explore interesting physics and develop quantum technologies through stacking different layers and forming heterostructures and moire structures \cite{he2021,mak2022}.  

\begin{figure*}
\center\includegraphics[scale=0.55]{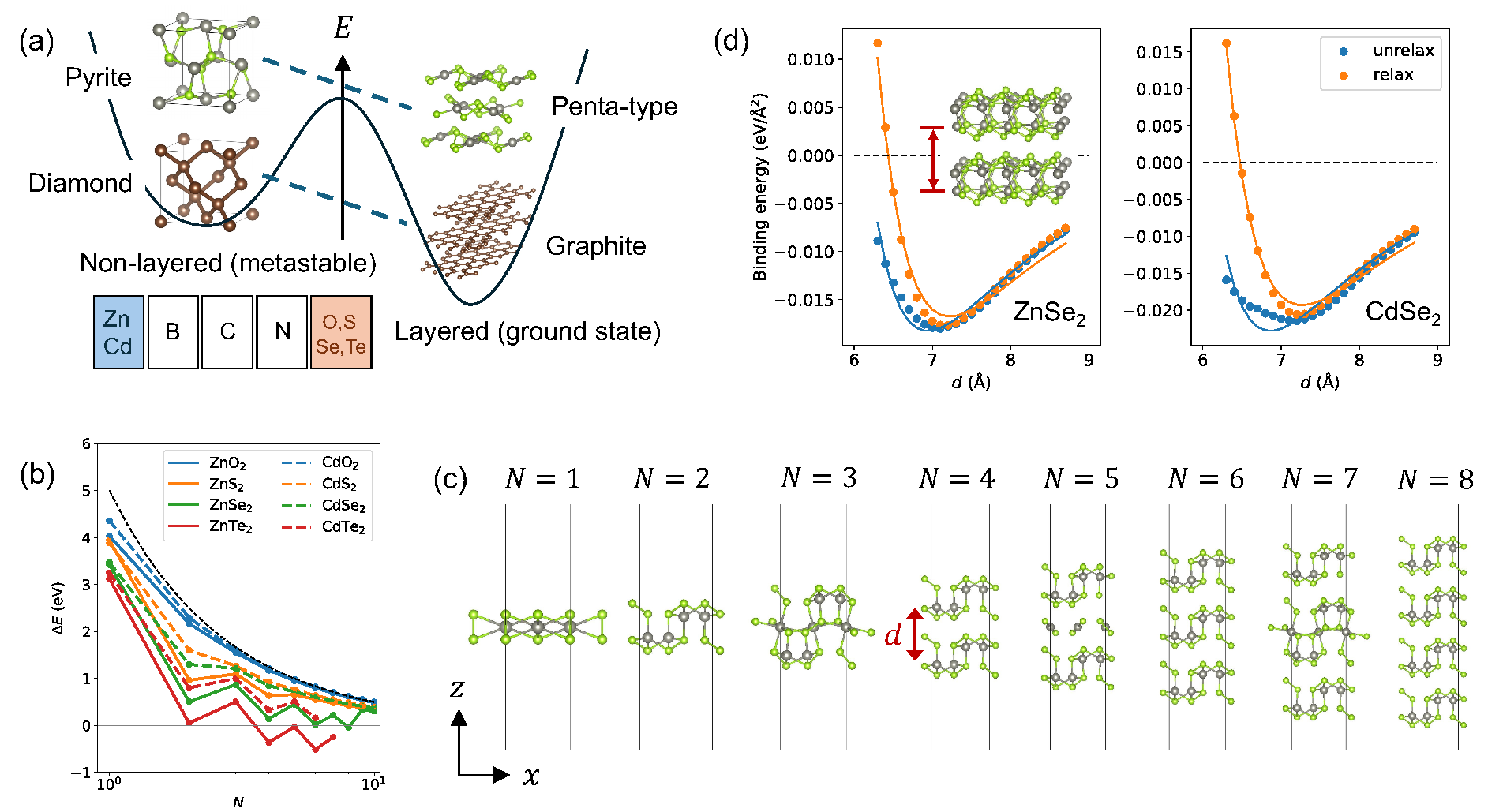}
\caption{(a) Schematic illustration of the potential energy surface. Carbon adopts a layered structure as the ground state and  a non-layered structure as a metastable state. II-VI semiconductors adopt the pyrite-type structure synthesized under high pressure conditions, in turn, implying the existence of a hidden layered structure. (b) $\Delta E$ as a function of $N$ for thin films derived from the pyrite (001) surface. The dashed line indicates $\Delta E \propto N^{-1}$. The slab model calculations were not converged for $N\ge 7$ in CdTe$_2$ and $N\ge 8$ in ZnTe$_2$, but the layered phase is preferred in the bulk limit (see Fig.~\ref{fig3}). (c) The side views of ZnSe$_2$ thin film structure up to $N=8$. $N=2$ thin film is terminated by Se atoms, serving as a building block for creating the vdW 2D ZnSe$_2$ with $N=4$-8. (d) The binding energy of $N=4$ ZnSe$_2$ and CdSe$_2$ as a function of the interlayer distance $d$ that is equal to the distance along the $z$ direction between Zn (Cd) atoms of different layers (see the inset). The rigid $N=2$ monolayer is assumed for ``unrelax'', and only the $z$ coordinates of Zn (Cd) pairs of different layers are fixed for ``relax''. The (12,6)-Lennard-Jones potential is used to fit the calculated data (filled circle). } \label{fig1} 
\end{figure*}

Graphene is a key ingredient in developing the 2D materials science, and can be exfoliated from graphite, where each layer is weakly bonded via the vdW forces \cite{graphene}. Carbon also adopts the diamond structure, which is categorized into non-vdW (i.e., non-layered) structure. The diamond structure is metastable compared to the layered structure, and such a structure has been synthesized under high-pressure conditions. These two allotropes have different chemical bonds (i.e., $sp^2$ and $sp^3$ bonds for graphite and diamond, respectively), and the phase stability between the diamond and graphite structures has been extensively studied for bulk \cite{khaliullin2011,popov} and ultrathin films \cite{lavini2022}. Is there a compound that, as for carbon, adopts both layered and non-layered structures? Is the existence of {\it metastable} non-layered structure a good indicator for the existence of {\it ground state} layered structure (see Fig.~\ref{fig1}(a))? 

Let us move on to the other group IV elements. Si, Ge, and Sn have the diamond structure as the ground state. Still, their monolayers derived from the (111) surface have buckled honeycomb structure \cite{sahin2009} and they are realized on several substrates \cite{vogt2012,takamura2012,reviewIV}. Although Pb has the face-centered cubic structure as the ground state, the Pb monolayer has also been created on metallic substrate \cite{yuhara}. 

Boron nitride (BN) is a wide-gap insulator, and adopts both layered (hexagonal) and non-layered (cubic) phases. 2D BN is exfoliated from the layered bulk. However, the ground state structure of BN is actually the non-layered phase having the zincblende-type structure---a binary derivative of the diamond structure \cite{halo2011,biswas}. This contrasts with carbon, where the ground state is the layered graphite. 



The III-V semiconductors, $AB$, have the zincblende-type structure. When truncating ultrathin films from the (111) surface, the double layer honeycomb structure becomes highly stable, and serves as topological insulators \cite{lucking2018,qin2021}. The stability of the double layer structure originates from the displacements of $A$ and $B$ atoms to form the ionic bonds between upper and lower layers. However, the stability of the zincblende-type structure overcomes the double layer honeycomb structure in the bulk limit. 

We next consider the II-VI semiconductors. While, as for the III-V semiconductors, they form the zincblend-type structure, they also have the pyrite-type (FeS$_2$-type) structure with a cubic symmetry (space group of $Pa\bar{3}$) and 12 atoms in the unit cell (see Fig.~\ref{fig1}(a)). $A$ atoms form the face-centered cubic lattice and each $A$ atom is surrounded by six $B$ atoms. More interestingly, the pyrite-type metal dichalcogenides ZnS$_2$, ZnSe$_2$, CdS$_2$, and CdSe$_2$ have been synthesized at 6.5-8.9 GPa \cite{bither}. Therefore, by analogy with the metastability relationship between layered and non-layered structures in carbon, we hypothesize that II-VI semiconductors should have a layered ground state, with the pyrite phase being metastable, as shown in Fig.~\ref{fig1}(a). The thin films derived from the (001) surface corresponds to the pentagon-shaped monolayer, and the chalcogen $B$ atoms are expected to terminate the thin film surface, yielding novel 2D materials, as for PdSe$_2$ \cite{oyedele2017}.

Note that the metal dichalcogenides mentioned above are different from the well-known transition metal dichalcogenides (TMD). They form a compound $AB_2$ with $A$ being group 4-10 metals, and most of 2D TMD adopts 2H, 1T, and 1T$^\prime$ structures \cite{TMD,silva2022}. 

Recently, the author has developed a framework for identifying non-vdW 2D materials \cite{ono2025}. This is based on the energy calculations of $N$-layer-thick thin films. The deviation from the $N^{-1}$ scaling of the thin film energy is a good descriptor for predicting non-vdW 2D materials. This framework successfully identifies known 2D materials such as group IV elements \cite{reviewIV} and goldene \cite{kashiwaya}. Application to carbon diamond also identifies the 3D-2D phase transition: when $N\le 4$, the thin films derived from the (111) surface relaxed to the $N$-layer graphene. Therefore, this framework can be used to explore a possible layered ground state of II-VI compounds.  

In this Letter, we demonstrate that in the ultrathin limit (less than 26 \AA), the pyrite-type ZnSe$_2$ spontaneously transforms into the layered phase. Such a layered structure is actually identified as a ground state at zero pressure. The layered structure consists of $N=2$ thin films terminated by chalcogen atoms (we call this as $N=2$ monolayer), and the ultrathin film is stacked along the surface normal via the vdW forces. The bandgap of $N=2$ monolayer is 1.71 eV (indirect) and decreases with the number of layers, whereas their magnitude is larger than of the pyrite phase by a factor of two. A two-valued but directional potential energy surface (PES) exists along the Bain-like distortion path, hiding the layered phase, in stark contrast to carbon. This work suggests that the metastability relationship between the layered and non-layered structures is useful for exploring novel layered materials, and also bridges 2D materials and surface and high-pressure science.


\begin{figure*}
\center\includegraphics[scale=0.45]{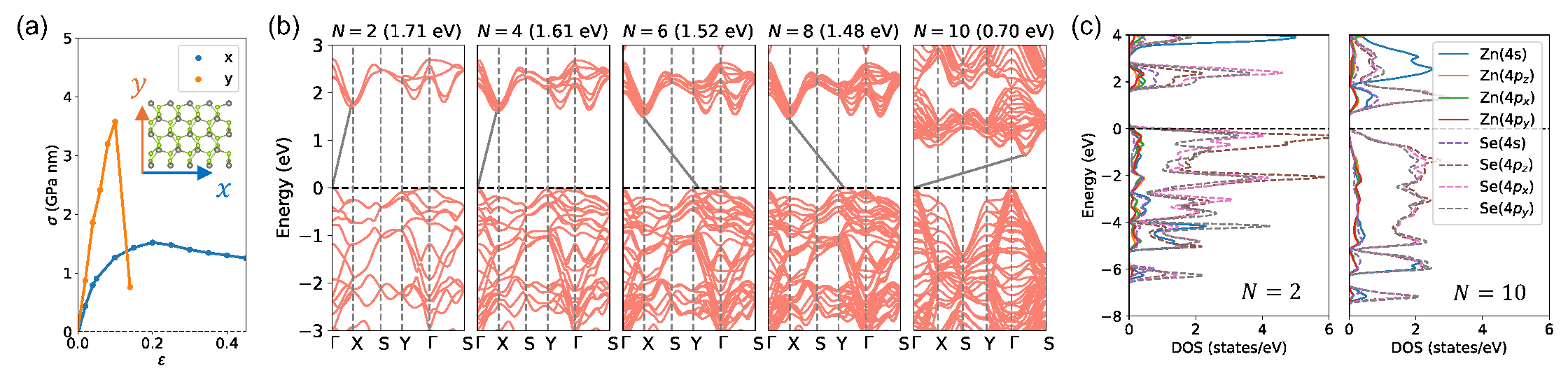}
\caption{(a) Stress-strain curves of $N=2$ ZnSe$_2$ monolayer. (b) Evolution of the electron band structure for $N$ ZnSe$_2$ thin films (PBE-GGA). The band gap at $N=10$ is small because the thin film keeps the initial pyrite-type geometry. (c) The projected density-of-states of $N=2$ and 10 thin films. Energy is measured from the Fermi level.  } \label{fig2} 
\end{figure*}

{\it Methods---}To identify the 3D-2D phase transition of II-VI semiconductors in the ultrathin limit, we calculated the finite-thickness excess energy defined as
\begin{eqnarray}
\Delta E (N) = \frac{E (N)}{N}-E_{\rm bulk},
\label{eq:FTEE}
\end{eqnarray}
where $E (N)$ and $E_{\rm bulk}$ are the total energy per formula unit of $N$-layer thin film and bulk, respectively. One can show that $\Delta E \propto N^{-1}$ for large $N$ by decomposing the total energy into a sum of the energy of each layer \cite{ono2025}. The $N$-layered thin films having $6N$ atoms in the unit cell are derived from the (001) surface of pyrite-type structure. If the layered ground state exists, a strong deviation from the $\Delta E \propto N^{-1}$ relation and negative $\Delta E$ will be observed. The geometry optimization of thin films was performed within density-functional theory (DFT) using Quantum ESPRESSO \cite{qe}. The PBE-GGA exchange-correlation functional \cite{pbe} with the Grimme's D3 correction was used \cite{dftd3}. The computational details are provided in Supplemental Material \cite{SM}. 

We considered $AB_2$ with $A=$ Zn and Cd and $B=$ O, S, Se, and Te. The synthesis of the pyrite-type phase was reported for oxides, sulfides, and selenides \cite{oxides,bither}. We also investigated $A$Te$_2$ as a reference.  

{\it Layered bulk and monolayer---}Figure \ref{fig1}(b) shows the $N$-dependence of $\Delta E$. The $N^{-1}$ scaling is clearly observed for the oxides $A$O$_2$ for large $N$. For other systems, distinct dip structures are found at even $N$, and the strength becomes strong as $B$ goes from S to Se to Te. It is emphasized that $\Delta E$ becomes negative at $N=8$ for ZnSe$_2$ and $N\ge 4$ for ZnTe$_2$, indicating that their thin films are more stable than the pyrite phase. 



Figure \ref{fig1}(c) shows the evolution of the thin film structure for ZnSe$_2$. The $N=2$ monolayer is realized by the Se atom movement toward outermost surfaces. Interestingly, ZnSe$_2$ thin films at $N=4$, 6, and 8 can be regarded as bi-, tri-, and quad-layers of $N=2$ ZnSe$_2$ monolayer, respectively. For example, for $N=4$, the interlayer distance is $d=7.04$ \AA \ that is larger than the intralayer Zn-Zn distance of 2.17 \AA \ along the $z$ axis. The lattice constant $a$ decreases from 7.211 \AA \ at $N=2$ to 7.141 \AA \ at $N=8$, whereas the ratio of $b/a$ increases from 0.769 at $N=2$ to 0.785 at $N=8$. 

These results lead us to propose that ZnSe$_2$ is layered, rather than pyrite-type material \cite{bither}. Through DFT calculations, we have confirmed that the layered bulk (space group of $Pca2_1$) is more stable than the pyrite phase by 0.02 eV/atom, and that no imaginary phonon frequencies are found in the entire Brillouin zone. The energy ordering of these structures is the same if we include the zero-point vibrational energy contribution, and this is also true for finite temperature within the quasi-harmonic approximation. We also confirmed that $N=2$ ZnSe$_2$ monolayer is dynamically stable at zero and finite temperatures. The exfoliation energy is estimated to be 19 meV/\AA$^2$, which is similar to that of typical vdW 2D materials \cite{park}. The dynamical stability of layered ZnTe$_2$ and CdTe$_2$ were also confirmed. These results are provided in Supplemental Material \cite{SM}. 

To study the vdW nature of ZnSe$_2$, we calculated the binding energy $E_{\rm B}$ of ZnSe$_2$ and CdSe$_2$ thin films at $N=4$, as shown in Fig.~\ref{fig1}(d). $E_{\rm B}$ of ZnSe$_2$ is well described by the (12,6)-Lennard-Jones (LJ) potential. On the other hand, $E_{\rm B}$ of CdSe$_2$ with the atom positions relaxed (except for Cd pairs) significantly deviates from the LJ curve for $d<7$ \AA, implying that the ionic and covalent bonds may contribute to the interlayer interactions.

Due to the puckered geometry along the $x$ direction in $N=2$ ZnSe$_2$ monolayer, the response of a tensile strain $\varepsilon$ exhibits strong anisotropy, as shown in Fig.~\ref{fig2}(a). The maximum stress is 1.5 and 3.6 GPa nm at $\varepsilon = 0.2$ and 0.1 along $x$ and $y$ directions, respectively. The anisotropy of the strain limits is larger than of phosphorene with $\varepsilon=0.27$ and 0.3 along the zigzag and armchair directions, respectively \cite{peng2014}. 

Figure \ref{fig2}(b) shows the evolution of the electronic structure for ZnSe$_2$ with $N=2, 4, 6, 8,$ and 10. The layered structures up to $N=8$ have an indirect bandgap that decreases from 1.71 eV at $N=2$ to 1.55 eV at $N=8$ (PBE-GGA). This is larger than that of $N=10$ pyrite-type thin film by a factor of two. When switching from GGA to HSE functional \cite{hse}, the bandgap for the $N=2$ monolayer is increased to 2.63 eV (see Supplemental Material \cite{SM}). The projected density-of-states reflects the 3D-2D transition (see Fig.~\ref{fig2}(c)): The edge of the valence and conduction bands mainly consists of $4p$ states of Se, and the contributions from the $4p_x$, $4p_y$, and $4p_z$ states are almost the same for the $N=10$ thin film, while those of $4p_z$ states are enhanced for the $N=2$ ZnSe$_2$ monolayer. The contribution from the Zn states also increases near the valence band maximum. 

\begin{figure*}
\center\includegraphics[scale=0.4]{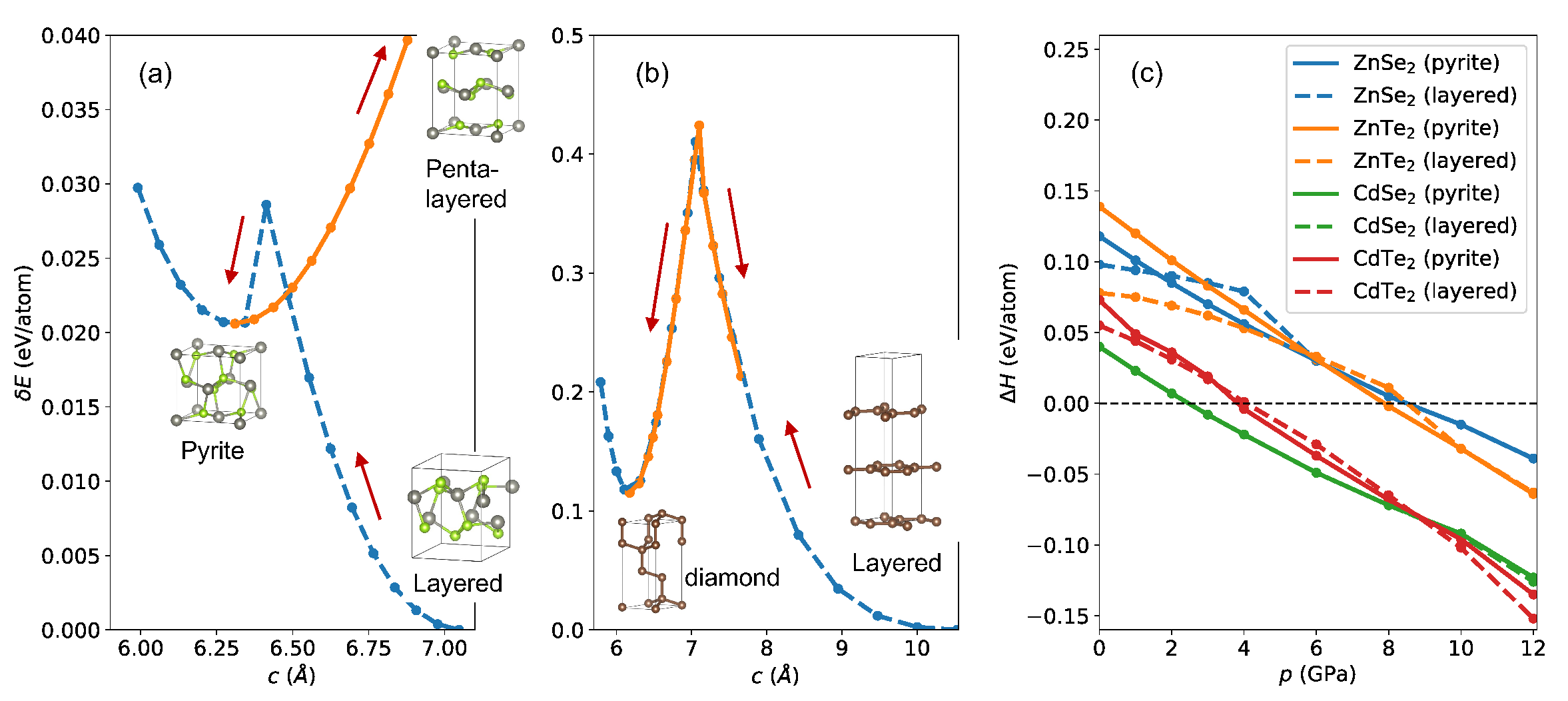}
\caption{Bain-like distortion, with $a$ and $b$ axes relaxed for each $c$, staring from the layered ground state structure for (a) ZnSe$_2$ and (b) C. Once ZnSe$_2$ forms the pyrite-type phase, the strain energy $\delta E$ increases with $c$ through different energy path leading to the orthorhombic phase like PdSe$_2$ \cite{oyedele2017}. On the other hand, the layered phase recovers if a tensile strain along the [111] direction is applied to the diamond carbon. (c) $\Delta H$ in Eq.~(\ref{eq:enthalpy}) of pyrite-type and layered phases. For CdSe$_2$, the layered structure relaxed to the pyrite-type phase.  } \label{fig3} 
\end{figure*}

{\it Hidden phase---}In the geometry optimization calculations for the $N$ layer thin films, we have prepared the initial geometry by cutting the thin films from the pyrite (001) surface. One can observe positive $\Delta E$ at $N=10$ for ZnSe$_2$ (see Fig.~\ref{fig1}(b)). This means that once the pyrite phase is formed, such a phase will be kept unless $N\le 8$ (i.e., quad-layers of $N=2$ monolayer). This is quite similar to the carbon thin films, where the diamond structure is kept unless $N\le 4$ \cite{ono2025}. 

To better understand the transformation mechanism between two phases in ZnSe$_2$, we performed the Bain-like distortion calculations \cite{grimvall2012} by starting from the layered ground state, with the length of $c$ fixed and that of $a$ and $b$ relaxed. As shown in Fig.~\ref{fig3}(a), the strain energy $\delta E$ increases with decreasing $c$, and at a critical $c\simeq 6.3$ \AA, $\delta E$ exhibits a significant decrease, accompanying the layered-to-pyrite phase transition. Conversely, $\delta E$ increases again with increasing $c$ by staring from the pyrite-type phase. Although such a bulk for large $c$ has the layered structure, they consists of the pentagon-shaped monolayers as of PdSe$_2$ \cite{oyedele2017}. This suggests that there are two branches in the PES, depending on the initial geometry, and also implies that the energy valley corresponding to the layered phase is hidden in the PES. This is qualitatively different from the reversible diamond-to-graphite transformation shown in Fig.~\ref{fig3}(b). Below and above a critical value of $c=7.1$ \AA, the geometry is relaxed to the diamond and graphite structures, respectively.  

The pyrite-type ZnSe$_2$ has been synthesized under high-pressure conditions \cite{bither}. To investigate the stability of the layered phase under pressure $p$, we calculate the formation enthalpy $\Delta H$ defined as 
\begin{eqnarray}
 \Delta H = H(AB_2) - H_{\rm ZB}(AB) - H_{\rm chal}(B),  
 \label{eq:enthalpy}
\end{eqnarray}
where $H(AB_2)$, $H_{\rm ZB}(AB)$, and $H_{\rm chal}(B)$ are the enthalpy per atom of $AB_2$ in the pyrite-type or layered structure, $AB$ in the zincblende-type structure, and $B$ in the trigonal structure. Figure \ref{fig3}(c) shows $\Delta H$ as a function of $p$ for selenides and tellurides. The layered structure is more stable than the pyrite-type structure at low $p$. $\Delta H$ decreases with increasing $p$. The stability of the pyrite phase overcomes the layered phase, and $\Delta H$ becomes negative at high $p$. 


We have also calculated $\Delta E(N)$ for other pyrite-type materials including 20 chalcogenides and 10 pnictides and PdF$_2$. The crystal structure of these materials are extracted from the Materials Project database \cite{materalsproject}. We have confirmed that no thin films overcome the stability of the pyrite bulk phase. Although a small dip at $N=2$ was found for SiP$_2$, no anomalies were observed for larger $N$. Therefore, ZnSe$_2$ is a peculiar system with the layered ground state. The calculated data for $\Delta E(N)$ are provided in Supplemental Material \cite{SM}. 

{\it Conclusions and outlook---}To investigate a hypothesis that a layered ground state phase should exist if non-layered structure having the same stoichometry exists as a metastable structure, we have explored a hidden layered phase for the metal dichalcogenides having the pyrite-type structure. This is motivated by an analogy of metastability in carbon, where the non-layered metastable structure is synthesized under high-pressure conditions. Through the energy variation calculations of $N$-layered thin films, we have identified a possible layered phase of ZnSe$_2$ that is more stable than the pyrite-phase at a low pressure condition. ZnSe$_2$ monolayer exhibits a strong elastic anisotropy and a semiconducting bandgap that is larger than of the pyrite-type phase by a factor of two. ZnTe$_2$ and CdTe$_2$ also adopt the same layered structure, whereas no pyrite-type phase has been reported. 

Through thin film calculations. we have demonstrated that the surface effect is significant to produce the layered structure, and the pyrite-type thin films with thickness less than 26 \AA \ (at $N=8$) spontaneously transform into the layered phase. The pyrite-type phase appears under a compressive strain along the $c$ axis, but the layered ground state does not recover even if a tensile strain applied to the pyrite phase. This suggests that a two-valued PES exists along a Bain-like distortion path, which hides the layered ground state. Although the layered ZnSe$_2$ may be prepared with well controlled conditions, the implication of the present work is that many layered materials, providing vdW 2D materials, are hidden in non-vdW (non-layered) 3D materials that have been synthesized under high pressure conditions. This connects 2D materials with both surface and high-pressure science. We hope that layered ZnSe$_2$, as a carbon-analog in a binary compound system, will be synthesized in future experiments. 


\begin{acknowledgments}
The author thanks Prof. S. Ebisu for useful discussion. This work was supported by JSPS KAKENHI (Grant No. 24K01142). Calculations were done using the facilities of the Supercomputer Center, the Institute for Solid State Physics, the University of Tokyo.
\end{acknowledgments}

\section*{Data availability}
The data that support the findings of this article are available from the corresponding author upon reasonable request.



\end{document}